\documentclass{aastex63}

\received{}
\revised{}
\accepted{}
\submitjournal{ApJ}

\shorttitle{Escape and accretion by cratering impacts}
\shortauthors{Hyodo \& Genda}

\graphicspath{{./figs/}}

\begin{document}

\title{Escape and accretion by cratering impacts: Formulation of scaling relations for high-speed ejecta}

\correspondingauthor{Ryuki Hyodo}
\email{ryuki.h0525@gmail.com}

\author[0000-0003-4590-0988]{Ryuki Hyodo}
\affiliation{ISAS, JAXA, Sagamihara, Japan}

\author[0000-0001-6702-0872]{Hidenori Genda}
\affiliation{Earth-Life Science Institute, Tokyo Institute of Technology, Tokyo 152-8550, Japan}

\begin{abstract}
Numerous small bodies inevitably lead to cratering impacts on large planetary bodies during planet formation and evolution. As a consequence of these small impacts, a fraction of the target material escapes from the gravity of the large body, and a fraction of the impactor material accretes onto the target surface, depending on the impact velocities and angles. Here, we study the mass of the high-speed ejecta that escapes from the target gravity by cratering impacts when material strength is neglected. We perform a large number of cratering impact simulations onto a planar rocky target using the smoothed particle hydrodynamics method. We show that the escape mass of the target material obtained from our numerical simulations agrees with the prediction of a scaling law under a point-source assumption when $v_{\rm imp} \gtrsim 12 v_{\rm esc}$, where $v_{\rm imp}$ is the impact velocity and $v_{\rm esc}$ is the escape velocity of the target. However, we find that the point-source scaling law overestimates the escape mass up to a factor of $\sim 70$, depending on the impact angle, when $v_{\rm imp} \lesssim 12  v_{\rm esc}$. Using data obtained from numerical simulations, we derive a new scaling law for the escape mass of the target material for $v_{\rm imp} \lesssim 12  v_{\rm esc}$. We also derive a scaling law that predicts the accretion mass of the impactor material onto the target surface upon cratering impacts by numerically evaluating the escape mass of the impactor material. Our newly derived scaling laws are useful for predicting the escape mass of the target material and the accretion mass of the impactor material for a variety of cratering impacts that would occur on large planetary bodies during planet formation.
\end{abstract}


\section{Introduction} \label{sec:intro}
Planetary bodies grow from small to large \citep[e.g.,][]{Saf72,Hay85}. The number of bodies becomes smaller as the size increases. This indicates that small impacts on a larger body are inevitable and much more frequent than collisions between similar-sized bodies. Collisions between similar-sized bodies are often characterized by a catastrophic process and have been extensively studied in terms of a specific energy $Q_{\rm D}^*$ required for a 50\% mass loss in the gravity regime \citep[e.g.,][]{Ben99,Asp10,Lei12,Jut15,Gen15,Mov16}.\\

In contrast, cratering impacts $-$ impacts of small bodies on large planetary bodies $-$ eject target materials near the impact point and, depending on impact conditions, accrete impactor materials to the target surface. Cratering impacts have been extensively studied through laboratory experiments \citep[e.g.,][]{Fuj77,Fuj80,Har85,Nak92,Mic07,Tsu15}, numerical simulations, and theoretical approaches \citep[e.g.,][]{Oke77,Mel84,Mel89,Shu06,Art08,Sve11}. An analytical scaling law was derived under a point-source assumption \citep[e.g.,][]{Hou83,Hol87,Hol93,Hou11,Hol12} where the size of the impact influence, as characterized by the crater size, is assumed to be much larger than that of the impactor. The point-source scaling law predicts the ejecta mass as a function of its ejection velocity $v_{\rm eje}$. The point-source scaling law reproduces the results of laboratory experiments in which the studied ejection velocity is much smaller than the impact velocity (i.e., $v_{\rm eje} \ll v_{\rm imp}$; \cite{Hou11}). The point-source scaling law has also been used in the context of the mass escape from the target via cratering impacts, where a fraction of the target material near the impact point escapes from its gravity field. In these mass escape cases, the considered ejection velocity of the target material could be only moderately larger than the escape velocity of the impacted body (i.e., $v_{\rm eje} \gtrsim v_{\rm esc}$). In this case, the size of the impact influence is only moderately larger than that of the impactor, and the point-scaling law might not be appropriate (a failing was qualitatively noted by the originators of the point source solution; \cite{Hou11}). However, the conditions in which the point-source scaling law are valid are quantitatively unclear.\\

Here, we aim to evaluate high-speed ejecta with an ejection velocity larger than the escape velocity of the target (hereafter, the escape mass). We distinguish two sources of escape mass: target material and impactor material. The point-source solution implicitly considers only ejecta originating from the target (the target material) and omits the contribution of the escape mass originating from the impactor (the impactor material). In reality, a fraction of the impactor material also escapes upon the cratering impact. The impactor-originated material that accretes onto the target surface (the accretion mass of the impactor material, $m_{\rm imp,acc}$) is given by the mass balance as $m_{\rm imp,acc} = m_{\rm imp} - m_{\rm imp,esc}$, where $m_{\rm imp}$ is the mass of the impactor and $m_{\rm imp,esc}$ is the escape mass that originates from the impactor.\\

In this study, we perform an extensive number of cratering impact simulations onto a planar rocky target that cover a wide range of impact parameters. We independently study the high-speed ejecta that originates from the target or the impactor. Based on the results of the impact simulations, we derive a scaling law for the escape mass originating from the target by the cratering impacts. Also, by evaluating $m_{\rm imp,esc}$ from the numerical simulations and by considering the mass balance discussed above, we estimate $m_{\rm imp,acc}$ and derive a scaling law for the accretion mass originating from the impactor onto the target surface upon the cratering impact.\\

In Section \ref{sec_previous}, we summarize the previously derived scaling laws of the ejecta by cratering impacts. In Section \ref{sec_method}, we describe our numerical methods. In Section \ref{sec_target}, we present the numerical results of the ejecta that originates from the target (the target material) and derive new scaling laws for the escape mass of the target material. In Section \ref{sec_impactor}, we show the numerical results of the ejecta that originates from the impactor (the impactor material) and derive a new scaling law for the accretion mass of the impactor material onto the target surface. In Section \ref{sec_application}, we discuss the applications of our newly derived scaling laws for planet formation. In Section \ref{sec_summary}, we summarize our paper.\\

\section{Previous scaling laws of cratering impacts}
\label{sec_previous}
As a consequence of the cratering impact $-$ an impact of a small body on a large target $-$, a fraction of the target material from the surface $M(>v_{\rm eje,tar})$ exceeds a given ejection velocity $v_{\rm eje,tar}$. Widely known pioneering studies \citep{Hou83,Hol87,Hol93,Hol07,Hou11} derived the $M(>v_{\rm eje,tar})-v_{\rm imp}$ relationship as a function of the impact velocity $v_{\rm imp}$ and impact angle $\theta$ under a point-source assumption, where the size of the impact crater is assumed to be much larger than that of the impactor. Using the point-source scaling law, the escape mass is expressed by setting $v_{\rm eje,tar}=v_{\rm esc}$, where $v_{\rm esc}$ is the escape velocity of the target, as follows (hereafter HH11):\\

\begin{equation}
\label{eq_HH11}
	\frac{M_{\rm HH11,esc,tar}(>v_{\rm esc})}{m_{\rm imp}} = C_{\rm HH11}\left( \frac{v_{\rm esc}}{v_{\rm imp}\sin \theta} \right)^{-3\mu_{\rm HH11}}
\end{equation}

\noindent where $C_{\rm HH11}=(3k/4\pi)C_{0}^{3\mu_{\rm HH11}}$ and $\mu_{\rm HH11}$ are constants. From the experiments, $\mu_{\rm HH11}=0.55$ for nonporous rocky and icy materials \citep[e.g.,][]{Hol07}. $k=0.3$ and $C_{\rm 0}=1.5$, respectively \citep{Gau63}. The densities of the impactor and target are assumed to be the same in the above equation.\\

Later, impact simulations of basaltic materials were performed by \cite{Sve11} to study the ejecta mass originating from the target for a wide range of impact parameters, $v_{\rm imp}=1.25-60$ km s$^{-1}$ and $\theta=0-90$ degrees. The results of the impact simulations are averaged over the statistical distribution of the impact angles of $\sin(2\theta)$ \citep{Sho62}. \cite{Sve11} reported that the $\theta$-averaged $M(>v_{\rm eje,tar})-v_{\rm imp}$ relationship has the same power-law dependence as \cite{Hou11} (Equation \ref{eq_HH11}), but with a different exponent for $v_{\rm imp} <10$ km s$^{-1}$. Ultimately, \cite{Sve11} derived a different ejection formulation by mathematically fitting the $\theta$-averaged numerical results as follows:

\begin{equation}
\label{eq_SV11}
	\left< \frac{M_{\rm SV11,tar,esc}(>v_{\rm esc})}{m_{\rm imp}} \right>_{\theta} =
	 \frac{ C_{\rm SV11}(v_{\rm imp}) \left( v_{\rm imp}^2 - K_{\rm SV11}(v_{\rm imp}) v_{\rm esc}^2 \right) }{v_{\rm esc}^2} \left( \frac{v_{\rm imp}}{20v_{\rm esc}} \right)^{R_{\rm SV11}(v_{\rm imp})}
\end{equation}

\noindent where $R_{\rm SV11}(v_{\rm imp})$, $C_{\rm SV11}(v_{\rm imp})$, and $K_{\rm SV11}(v_{\rm imp})$ depend on $v_{\rm imp}$.\\

These scaling laws have been widely used in many studies to estimate the escape mass of target materials for a given impact. However, the specific impact parameters in which the point-source assumption and Equation \ref{eq_HH11} are valid are unclear. Moreover, it is unclear from the $\theta$-averaged scaling law (Equation \ref{eq_SV11}) how the escape mass changes for different impact angles. Additionally, some of the coefficients of Equation \ref{eq_SV11} \citep{Sve11} only address specific impact velocities; thus, the users themselves must interpolate these limited values for an arbitrary impact velocity. Below, using numerical simulations, we derive a new scaling law for the escape mass of target material by cratering impacts. Our newly derived scaling law for the escape mass of the target material is combined with the point-source solution (HH11) and covers a wide range of impact parameters, even those beyond the limitation of the point-source assumption (see Equation \ref{eq_new_target}).\\

\section{Numerical methods of cratering impacts}
\label{sec_method}
\begin{figure*}
	\centering
	  \includegraphics[width=\textwidth]{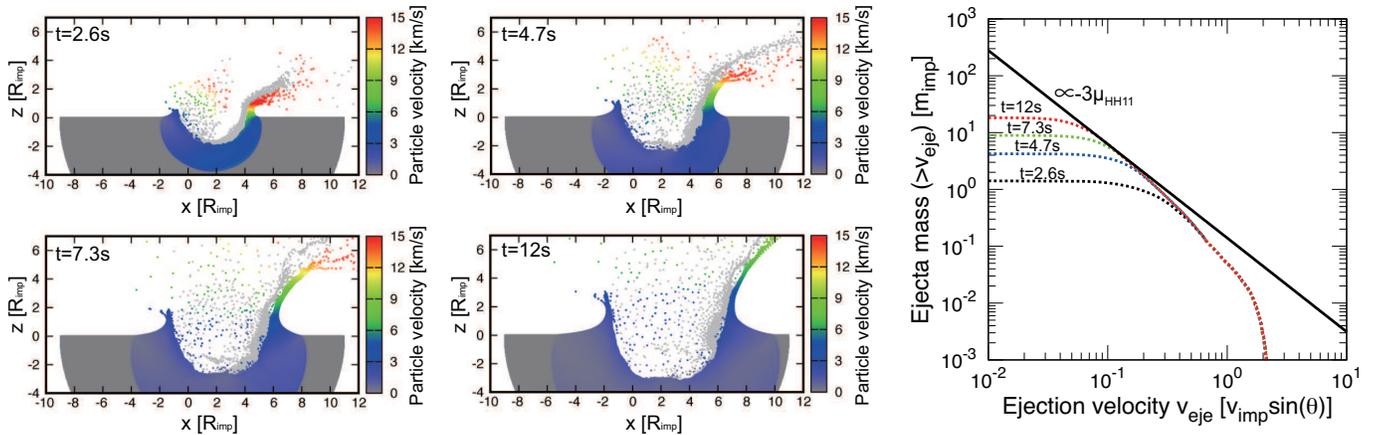}
	\caption{Left 4 panels: snapshots of our typical impact simulations at different epochs. Right panel: a time-evolution of the cumulative mass of the ejecta of the target material that exceeds a given ejection velocity. Included is the case of $v_{\rm imp}=27$ km s$^{-1}$ and $\theta=45$ degrees. Left 4 panels: $x$- and $z$-axes are in the unit of the impactor radius $R_{\rm imp}$. The vector of the impact velocity has $+ x$ direction in the $x$-$z$ plane. Only SPH particles that exist within $y = \pm 0.1R_{\rm imp}$ are plotted. The color contour indicates the particle velocity in a unit of km s$^{-1}$. Impactor materials are plotted in gray. $t = 0$ indicates when the impactor touches the surface of the target. Right panel: black, blue, green, and red dashed lines are the results of simulations at different epochs that correspond to the four left panels. The solid black line represents the point-source scaling law ($-3\mu_{\rm HH11}$; Equation \ref{eq_HH11}). The ejection velocity becomes larger for particles with launch points closer to the impact point (see four left panels). Thus, the numerical results gradually converge to a smaller ejection velocity.}
	\label{fig_snapshots}
\end{figure*}

In this study, we evaluate the high-speed ejecta that escapes from the gravity of the target $-$ the escape mass $-$ for a given impact velocity and impact angle using direct impact simulations of cratering impacts. We independently studied distinct sources of the escape mass: target material (Section \ref{sec_target}) and impactor material (Section \ref{sec_impactor}). We used the three-dimensional smooth particle hydrodynamic (SPH) method \citep{Luc77,Mon92} for cratering impacts on a planar target (Figure \ref{fig_snapshots}). Our numerical code was the same as that used in previous studies \citep{Gen15,Kur19,Hyo19}. We neglected gravity and material strength, and our numerical results are valid for high-speed ejecta where the material strength is negligible. The ejection behavior obtained from our SPH simulations was reproduced by the recent impact experiments \citep{Oka20}.\\

The impactor was represented by a spherical projectile with a radius of $R_{\rm imp}=10$ km. The target was represented by the flat surface of a half-sphere target with a radius of typically 10 times that of the projectile. For the numerical resolution, $3.4 \times 10^4$ SPH particles were used for the projectile, which corresponds to approximately 20 SPH particles per projectile radius (PPPR). Equal-mass SPH particles were used as the target, which corresponds to $1.7 \times 10^7$ SPH particles. To check the convergence of the ejection velocity distribution, we used a larger target with a radius 20 times that of the projectile. We used the Tillotson equation of state \citep{Til62} with the parameter sets for basalt \citep{Ben99} for both the projectile and the target. We confirmed that our numerical results converged by comparing 20 PPPR and 10 PPPR cases for the target escape. In the 10 PPPR cases, the impactor was represented by 4819 SPH particles, and the target was represented by $2.1 \times 10^6$ SPH particles.\\

For the impact parameters, we considered various impact velocities that typically ranged from 6 km s$^{-1}$ to 62 km s$^{-1}$ with a 7 km s$^{-1}$ interval and various impact angles from 15-90 degrees with a 15-degree interval, where a 90-degree impact is a head-on/vertical impact. For an extreme case, we performed additional calculations with $v_{\rm imp}=90$ km s$^{-1}$ and  $\theta=90$ degrees. We note that no data source from the experiments is available for such high-speed impacts (up to $90$ km s$^{-1}$); thus, such computations are hypothetical. The numerical setting of the planar target is valid for small impactors onto large planetary bodies $-$ the cratering impacts $-$ where the curvature of the target is negligible \citep{Gen17}. Although only one set was considered for the impactor radius, we were able to convert our results to any impactor size, because all hydrodynamic equations can be rewritten in a dimensionless form without gravity and strength. The densities of the impactor and target were equalized, and we will evaluate the dependence on the relative differences in densities between the target and the impactor in a future study.\\

\section{Escape of target material by cratering impacts}
\label{sec_target}
In this section, we discuss the escape mass that originates from the target (the target material). First, we explain the typical outcome of our numerical simulation (Section \ref{sec_typical}). Then, we investigate the ejection velocity distributions for different impact angles (Section \ref{sec_ejection}). Finally, we derive a new scaling law for the escape mass of the target materials that can be used for a wide range of impact parameters (Section \ref{sec_scaling_target}).\\

\subsection{A typical outcome from crater-forming impacts}
\label{sec_typical}
After the contact of the impactor with the target surface, the impact shock propagates through the interior of the target, and crater formation occurs as the material is sheared, moving upward and outward along the bowl-shaped crater edge \citep[Figure \ref{fig_snapshots}; see more details in][]{Hou11}. The distribution of the ejection velocity is related to its launch position \citep[e.g.,][]{Pie80} in a manner where high- and low-speed ejecta are launched closer to and farther from the impact point, respectively (Figure \ref{fig_snapshots}). When analyzing the numerical simulations, we defined the ejecta particles as those existing above the surface of the target. As a final snapshot of our simulations, $-$ and thus the data used for the analysis $-$ we used the snapshots when the shock reaches the boundary of the target.\\

As shown in Figure \ref{fig_snapshots}, the distribution of the ejection velocity obtained from our numerical simulation (dashed lines) gradually converges to smaller values, because the lower-speed ejecta is launched farther from the impact point with time. The result of the numerical simulation matches the point-source scaling law (solid black line in Figure \ref{fig_snapshots}; Equation \ref{eq_HH11}) below a certain value of ejection velocity (e.g., $ \sim 0.2 v_{\rm imp} \sin \theta$ in Figure \ref{fig_snapshots}). This indicates that the point-source assumption is valid for the low-speed ejecta regime. In contrast, the high-speed ejecta regime deviates from the point-source scaling law (e.g., $\gtrsim 0.2 v_{\rm imp} \sin \theta$ in Figure \ref{fig_snapshots}), as qualitatively expected by \cite{Hou11}. Our numerical settings, such as the size of the target and the simulation time, are chosen so that the ejection velocity distribution converges at a sufficiently small value to resolve the changes in these different regimes. We note that the point-source scaling law would not be appropriate for a very small ejection velocity because of the effects of strength and gravity \citep[see][]{Mel89,Hou11}, even though such a regime does not pertain to this study. In the next subsection, we present the results of a variety of impact parameters.\\

\subsection{Ejection velocity of target material}
\label{sec_ejection}
\begin{figure*}
	\centering
	  \includegraphics[width=\textwidth]{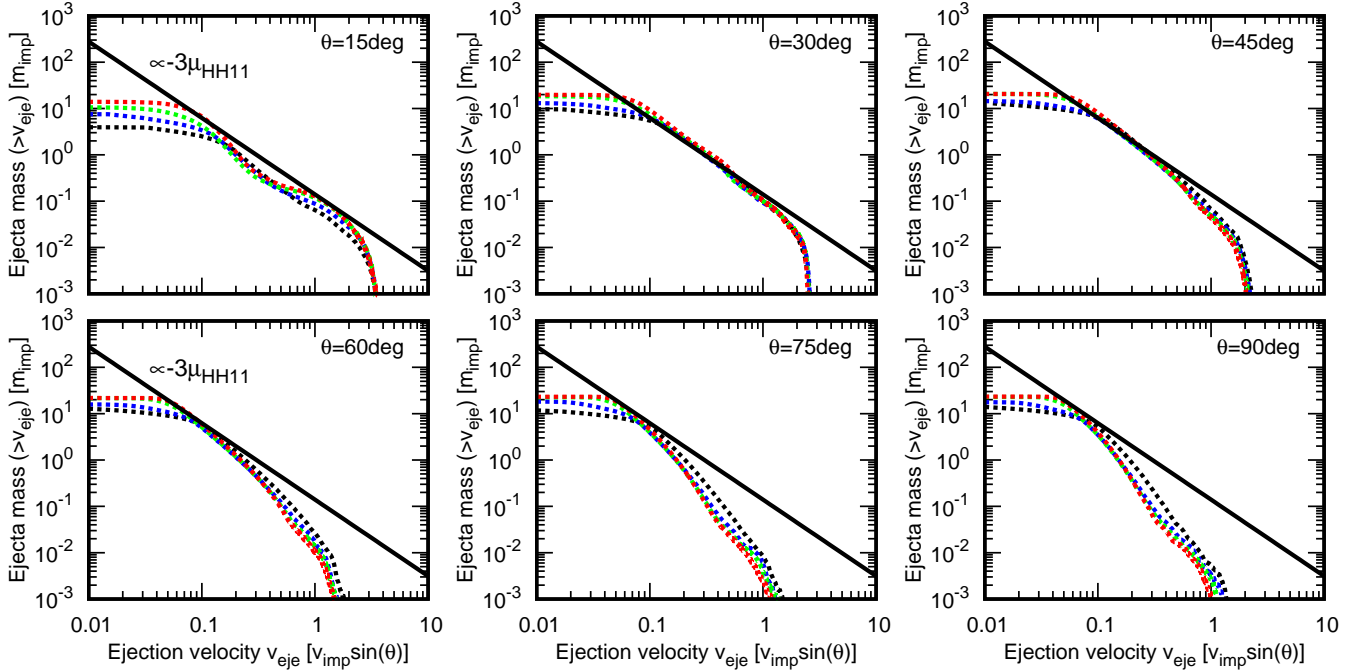}
	\caption{Cumulative mass ($>v_{\rm eje}$) of the impact ejecta that exceeds a given ejection velocity $v_{\rm eje}$ for a different impact angle $\theta$. Only target materials are plotted. Red, green, blue and black dashed lines are the results of SPH simulations for $v_{\rm imp}=62, 34, 13$ km s$^{-1}$ and $6$ km s$^{-1}$, respectively. The solid black line represents the point-source scaling law ($-3\mu_{\rm HH11}$; Equation \ref{eq_HH11}). Numerical results do not converge for small velocities where the point-source scaling assumption is valid, but this does not affect the conclusion of our study (see also Figure \ref{fig_snapshots}).}
	\label{fig_velocity}
\end{figure*}

Figure \ref{fig_velocity} shows the cumulative mass of ejecta originating from the target as a function of ejection velocity $v_{\rm eje}$ for different impact velocities and impact angles. The ejection velocity is scaled by $v_{\rm imp} \sin \theta$. The data for the ejection velocity with $v_{\rm eje} \lesssim 0.1 v_{\rm imp} \sin \theta$ does not converge (see the discussion in the previous subsection). The point-source scaling law (Equation \ref{eq_HH11}) is plotted as a solid black line. Figure \ref{fig_velocity} demonstrates the following important fundamentals of the ejection processes: (1) the distribution is uniquely scaled by $v_{\rm imp} \sin \theta$, which is in accordance with the spirit of the point-source assumption \citep{Hol07}, (2) the point-source scaling law is only valid for a limited range of ejection velocity distributions, and (3) the ejection velocity distribution has a unique power-law dependence for different impact angles; this dependence is not always the same as that of the point-source assumption ($-3\mu_{\rm HH11}$). Thus, a power law function can be used beyond the limitation of the point-source assumption by correcting its coefficient and exponent to depend on impact angles. On the one hand, the point-source assumption (Equation \ref{eq_HH11}) matches the numerical results at a small ejection velocity regime with launch points at large distances from the impact point. On the other hand, the slopes of the ejection velocity distributions are steeper than those of the point-source assumption ($-3\mu_{\rm HH11}$) for the high-speed ejection regime, where launch points are near the impact point. In the following subsection, we use these arguments to derive a new scaling law that includes the dependence on impact angles.\\

\subsection{Scaling law for escape mass of target material by cratering impacts}
\label{sec_scaling_target}
\begin{figure*}
	\centering
	  \includegraphics[width=\textwidth]{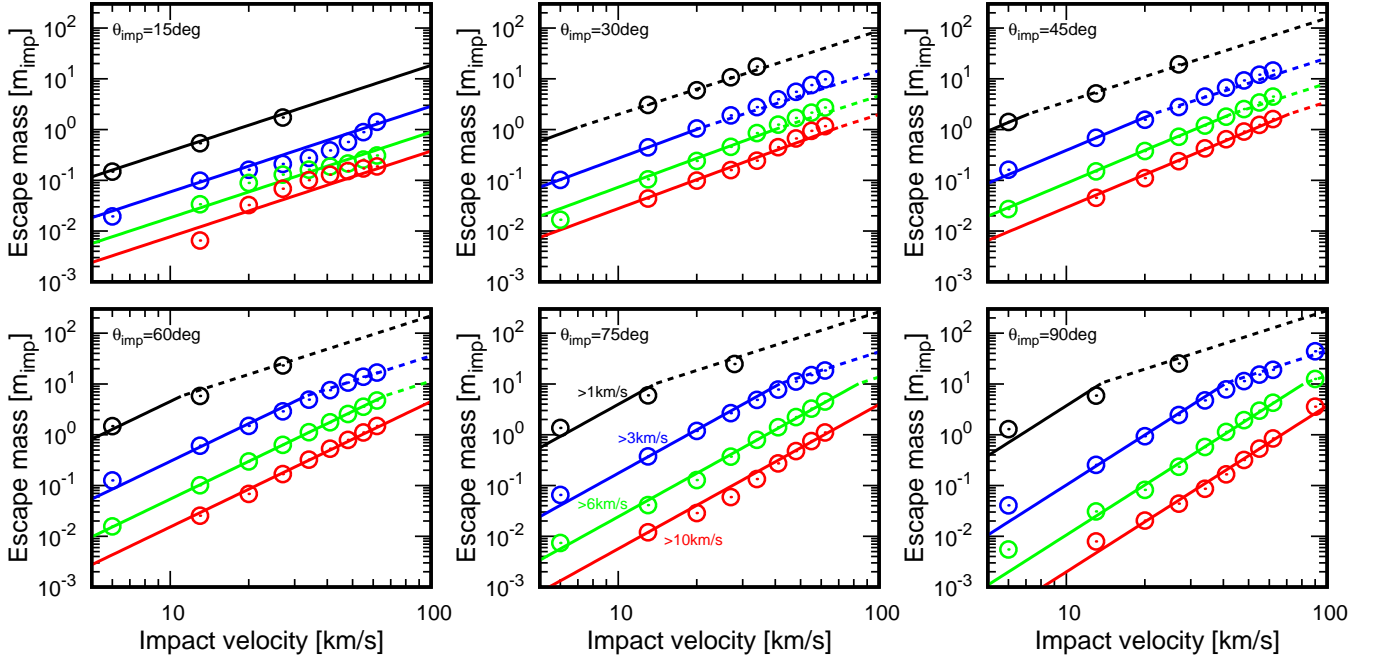}
	\caption{Escape mass of target material as a function of impact velocity for different impact angles. The escape mass of the target material is scaled by the mass of the impactor. Points are the results of the SPH simulations. Solid and dashed lines represent the new scaling law (Equation \ref{eq_new_target}; $\min \left\{ M^{*}_{\rm HG20,esc,tar}, M_{\rm HH11,esc,tar} \right \}$) derived in this work. Solid lines represent the cases where the same function as in Equation \ref{eq_HG20_target} ($M^{*}_{\rm HG20,esc,tar}$) is used. Dashed lines depict cases in which the function is the same as the Equation \ref{eq_HH11} ($M_{\rm HH11,esc,tar}$). Black, blue, green, and red lines represent cases where $v_{\rm esc}=1, 3, 6$ , and $10$ km s$^{-1}$, respectively.}
	\label{fig_target}
\end{figure*}

\begin{figure*}
	\centering
	  \includegraphics[width=0.8\textwidth]{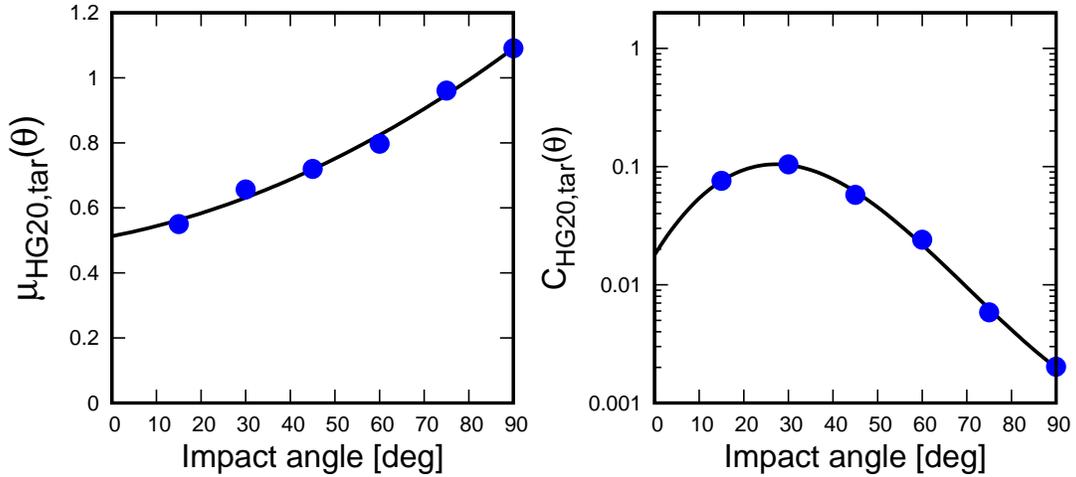}
	\caption{The exponent $\mu_{\rm HG20,tar}(\theta)$ (left) and the coefficient $C_{\rm HG20,tar}(\theta)$ (right) for the new scaling law of the escape mass of the target material (Equations \ref{eq_HG20_target} and \ref{eq_new_target}) as a function of the impact angle. Points are the results of SPH simulations and solid curves are the fitted quadratic and cubic functions of the impact angle for $\mu_{\rm HG20,tar}(\theta)$ and $C_{\rm HG20,tar}(\theta)$, respectively.}
	\label{fig_mu_C_target}
\end{figure*}

Figure \ref{fig_target} shows the escape mass of the target material whose ejection velocity is larger than a given value of escape velocity $v_{\rm esc}$ as a function of the impact velocity. Points were obtained from our numerical simulations. From the arguments in the previous subsection, we expect that the escape mass of the target material $M^{*}_{\rm HG20,esc,tar}$ for the high-speed ejecta regime ($\gtrsim 0.2 v_{\rm imp} \sin \theta$) is uniquely expressed by a power law as a function of impact velocity and impact angle as follows:

\begin{equation}
\label{eq_HG20_target}
	\frac{M^{*}_{\rm HG20,esc,tar}(>v_{\rm esc})}{m_{\rm imp}} =\\
	 C_{\rm HG20,tar}(\theta) \left( \frac{v_{\rm esc}}{v_{\rm imp}\sin \theta } \right)^{-3\mu_{\rm HG20,tar}(\theta)}
\end{equation}

\noindent where $C_{\rm HG20,tar}(\theta)$ and $\mu_{\rm HG20,tar}(\theta)$ are a new coefficient and a new exponent that depends on the impact angle, respectively. The physical meaning of $C_{\rm HG20,tar}(\theta)$ is the amount of ejecta. The physical meaning of $\mu_{\rm HG20,tar}(\theta)$ is the ejection velocity distribution among the ejecta. We fit the numerical results of SPH simulations (solid lines in Figure \ref{eq_HG20_target}) for the high-speed ejecta regime using Equation \ref{eq_HG20_target} and obtained $C_{\rm HG20,tar}(\theta)$ and $\mu_{\rm HG20,tar}(\theta)$ at different impact angles.  We then derived $\mu_{\rm HG20,tar}(\theta)$ and $C_{\rm HG20,tar}(\theta)$ (lines in Figure \ref{fig_mu_C_target}) using the quadratic and cubic functions of the impact angle, as follows:

\begin{equation}
\label{eq_mu_HG20}
	\mu_{\rm HG20,tar}(\theta) = a_{\rm tar}\theta^2 + b_{\rm tar}\theta + c_{\rm tar}
\end{equation}

\begin{equation}
\label{eq_C_HG20}
	C_{\rm HG20,tar}(\theta) = \exp \left( d_{\rm tar}\theta^3 + e_{\rm tar}\theta^2 + f_{\rm tar}\theta + g_{\rm tar} \right)
\end{equation}

\noindent where $a_{\rm tar}$, $b_{\rm tar}$, $c_{\rm tar}$, $d_{\rm tar}$, $e_{\rm tar}$, $f_{\rm tar}$ and $g_{\rm tar}$ are the fitted parameters, respectively (Table \ref{table_param}).\\

As discussed in the previous subsection, the original point-source scaling law (HH11; Equation \ref{eq_HH11}) is valid for sufficiently large distances to the impact point (i.e., for a sufficiently small ejection velocity). However, the coefficient and exponent deviate from those of HH11 (compare the solid black line and dashed lines in Figure \ref{fig_velocity}) for large ejection velocities or small distances to the impact point. As observed in Figure \ref{fig_velocity}, HH11 overestimates the escape mass of the target material, especially for a large ejection velocity. Therefore, the new scaling law for the escape mass of the target material that combined with HH11 is given by $\min \left\{ M^{*}_{\rm HG20,esc,tar}, M_{\rm HH11,esc,tar} \right \}$ and is written as follows (hereafter HG20):

\begin{equation}
\label{eq_new_target}
	\frac{M_{\rm HG20,esc,tar}(>v_{\rm esc})}{m_{\rm imp}} = 
	\min \left\{ C_{\rm HG20,tar}(\theta) \left( \frac{v_{\rm esc}}{v_{\rm imp}\sin(\theta)} \right)^{-3\mu_{\rm HG20,tar}(\theta)}, \right. 
 \left. C_{\rm HH11}\left( \frac{v_{\rm esc}}{v_{\rm imp}\sin(\theta)} \right)^{-3\mu_{\rm HH11}}  \right\}. 
\end{equation}
\\

\begin{table*}[]
\begin{center}
\begin{tabular}{|c|c|c|c|}
\hline

$a_{\rm tar}$         & $b_{\rm tar}$          & $c_{\rm tar}$         &               \\ \hline
$4.12 \times 10^{-5}$ & $2.71 \times 10^{-3}$  & $5.13 \times 10^{-1}$ &               \\ \hline
\hline
$d_{\rm tar}$         & $e_{\rm tar}$          & $f_{\rm tar}$         & $g_{\rm tar}$ \\ \hline
$1.52 \times 10^{-5}$ & $-3.21 \times 10^{-3}$ & $1.41 \times 10^{-1}$ & $-4.02$       \\ \hline

\hline
\hline

$a_{\rm imp}$         & $b_{\rm imp}$          & $c_{\rm imp}$         &               \\ \hline
$-7.06 \times 10^{-5}$ & $1.54 \times 10^{-2}$  & $-1.93 \times 10^{-1}$ &               \\ \hline
$d_{\rm imp}$         & $e_{\rm imp}$          & $f_{\rm imp}$         & $g_{\rm imp}$ \\ \hline
$3.34\times 10^{-5}$ & $-5.13 \times 10^{-3}$ & $1.28 \times 10^{-1}$ & $-8.71 \times 10^{-1}$       \\ \hline

\end{tabular}

\caption{Parameters of the fitted polynomial and exponential functions for the escape masses of the target material (Equation \ref{eq_new_target}) and the impactor material (Equation \ref{eq_HG20_impactor}), respectively.}
\label{table_param}
\end{center}
\end{table*}

In Figure \ref{fig_target}, the new scaling law (Equation \ref{eq_new_target}) is plotted by solid and dashed lines. We observed a close match between the new scaling law and impact simulations (points) for impact velocities of $6-62$ km s$^{-1}$ and impact angles of $15-90$ degrees for $v_{\rm esc}=1-10$ km s$^{-1}$.\\

\section{Accretion of impactor material by cratering impacts}
\label{sec_impactor}
In this section, we discuss the accretion mass of the impactor material onto the target surface for a variety of impact conditions at a variety of cratering impacts. A fraction of the impactor material escapes upon cratering impact, and the rest accretes onto the surface of the target. The accretion mass originating from the impactor $-$ the accretion mass of the impactor material, $m_{\rm imp,acc}$ $-$ is given by the mass balance as $m_{\rm imp,acc} = m_{\rm imp} - m_{\rm imp,esc}$, where $m_{\rm imp}$ is the mass of the impactor, and $m_{\rm imp,esc}$ is the escape mass originating from the impactor. Our SPH simulations without strength onto a planar target can resolve the high-speed ejecta, that is $m_{\rm imp,esc}$. By evaluating $m_{\rm imp,esc}$ from SPH simulations and by considering the mass balance, we estimate $m_{\rm imp,acc}$ and derive a scaling law of the accretion mass of the impactor material onto the target surface as shown below.\\

Figure \ref{fig_impactor} shows the mass of an impactor whose ejection velocity is larger than $v_{\rm esc}$; we define the escape mass of the impactor material as $M_{\rm HG20,esc,imp}(>v_{\rm esc})$. Points represent data obtained from the SPH simulations. Some of our simulations for high impact velocities and impact angles did not converge within a reasonable computational time. This capability was beyond the current computational resources available to us, and we will complete the calculations of these parameters in a future study. Such unconverged cases are plotted using triangles. Conversely, we confirmed the numerical convergence for the parameters plotted by circles in Figure \ref{fig_impactor}.\\

Using the same arguments as the target (Section \ref{sec_scaling_target}), we assume that $M_{\rm HG20,esc,imp}(>v_{\rm esc})$ follows a power law function as follows:

\begin{equation}
\label{eq_HG20_impactor}
	\frac{M_{\rm HG20,esc,imp}(>v_{\rm esc})}{m_{\rm imp}} = 
	 C_{\rm HG20,imp}(\theta) \left( \frac{v_{\rm esc}}{v_{\rm imp}\sin(\theta)} \right)^{-3\mu_{\rm HG20,imp}(\theta)}
\end{equation}

\noindent where $C_{\rm HG20,imp}(\theta)$ and $\mu_{\rm HG20,imp}(\theta)$ are coefficients and exponents that depend on the impact angle. We fit Equation \ref{eq_HG20_impactor} to our converged numerical results at $v_{\rm esc}=10$ km s$^{-1}$ to obtain the coefficient and the exponent at different impact angles.\\

\begin{figure*}
	\centering
	  \includegraphics[width=0.8\textwidth]{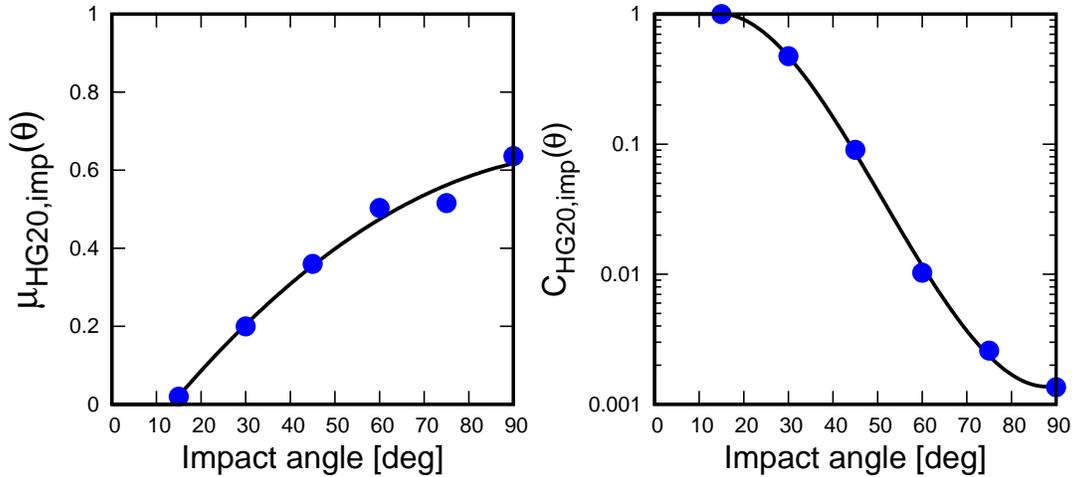}
	\caption{The exponent $\mu_{\rm HG20,imp}(\theta)$ (left) and the coefficient $C_{\rm HG20,imp}(\theta)$ (right) for the new scaling law (Equation \ref{eq_HG20_impactor}) that predicts the escape mass of the impactor material as a function of the impact angle. Points are the results of SPH simulations and solid curves are the fitted quadratic and cubic functions of impact angle for $\mu_{\rm HG20,imp}(\theta)$ and $C_{\rm HG20,imp}(\theta)$, respectively. Note that, $\mu_{\rm HG20,imp}(\theta) = 0$ and $C_{\rm HG20,imp}(\theta) = 1$ for $\theta < 15$ degrees.}
	\label{fig_mu_C_impactor}
\end{figure*}

The coefficients and exponents obtained from the numerical simulations are shown in Figure \ref{fig_mu_C_impactor}. As was performed for the target $-$ using the quadratic and cubic functions of the impact angle $-$ we derived $\mu_{\rm HG20, imp}(\theta)$ and $C_{\rm HG20, imp}(\theta)$ (lines in Figure \ref{fig_mu_C_impactor}), respectively, as follows: 

\begin{equation}
\label{eq_mu_HG20}
	\mu_{\rm HG20,imp}(\theta) = a_{\rm imp}\theta^2 + b_{\rm imp}\theta + c_{\rm imp}
\end{equation}

\begin{equation}
\label{eq_C_HG20}
	C_{\rm HG20,imp}(\theta) = \exp \left( d_{\rm imp}\theta^3 + e_{\rm imp}\theta^2 + f_{\rm imp}\theta + g_{\rm imp} \right)
\end{equation}

\noindent where $a_{\rm imp}$, $b_{\rm imp}$, $c_{\rm imp}$, $d_{\rm imp}$, $e_{\rm imp}$, $f_{\rm imp}$, and $g_{\rm imp}$ are the fitted parameters, respectively (Table \ref{table_param}). Note that, $\mu_{\rm HG20,imp}(\theta) = 0$ and $C_{\rm HG20,imp}(\theta) = 1$ for $\theta < 15$ degrees.\\

Our scaling law (Equation \ref{eq_HG20_impactor}) is plotted in Figure \ref{fig_impactor} by solid lines. Our newly derived scaling law of the escape mass of the impactor material generally agrees with the numerical results (points in Figure \ref{fig_impactor}), especially for $\theta = 30-75$ degrees. In the case of $\theta=90$ degrees, the scaling law deviates from the numerical results as the impact velocity increases. However, impacts with $\theta=90$ degrees do not statistically occur in planet formation, because the impact angle distribution is $\sin(2\theta)$ \citep{Sho62}.\\

Therefore, the accretion mass of the impactor material onto the target surface by cratering impacts is written by considering the mass balance and using Equation \ref{eq_HG20_impactor} as follows:

\begin{equation}
\label{eq_accretion}
	\frac{M_{\rm HG20,acc,imp}(<v_{\rm esc})}{m_{\rm imp}} = 
	 1 - C_{\rm HG20,imp}(\theta) \left( \frac{v_{\rm esc}}{v_{\rm imp}\sin \theta} \right)^{-3\mu_{\rm HG20,imp}(\theta)}.
\end{equation}
\\

\begin{figure*}
	\centering
	  \includegraphics[width=\textwidth]{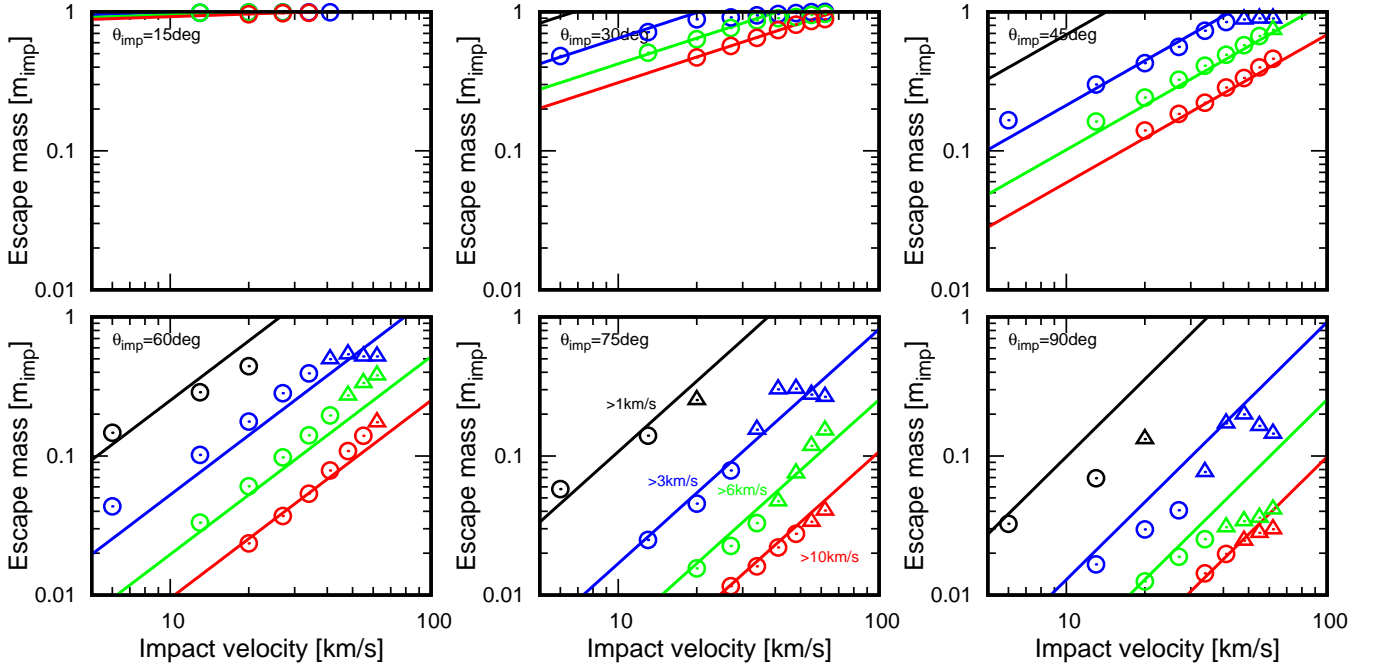}
	\caption{Escape mass of the impactor material as a function of impact velocity for different impact angles. The escape mass is scaled by the mass of the impactor. Points are the results of the SPH simulations. Triangles are used when numerical simulations do not converge. Circles are employed when numerical simulations converge. Solid lines represent the new scaling law (Equation \ref{eq_HG20_impactor}) derived in this study. Black, blue, green, and red lines represent the cases of $v_{\rm esc}=1, 3, 6$, and $10$ km s$^{-1}$, respectively.}
	\label{fig_impactor}
\end{figure*}

\section{Applications for planet formation}
\label{sec_application}
\subsection{Comparison between our study and the point-source scaling law}
\begin{figure}
	\centering
		\includegraphics[scale=0.8]{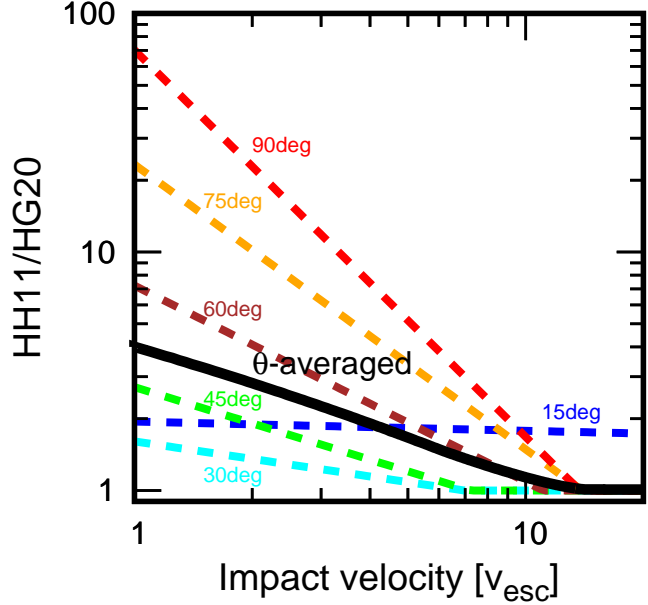}
	\caption{Ratio of the escape mass of the target material predicted by the point-source scaling law (HH11; Equation \ref{eq_HH11}) to that predicted by the newly derived scaling law (HG20; Equation \ref{eq_new_target}) as a function of the impact velocity in a unit of escape velocity of a target. Red, orange, brown, green, cyan and blue lines are the cases of the impact angles of $\theta=$90, 75, 60, 45, 30 and 15 degrees, respectively. The solid black line is the $\theta$-averaged case over $\sin(2\theta)$.}
	\label{fig_comparison}
\end{figure}

The point-source scaling law (HH11; Equation \ref{eq_HH11}) has been widely used in many studies. As shown in Section \ref{sec_target}, the prediction of the escape mass of the target material by the point-source scaling law agrees with the numerical results when $v_{\rm imp} \gg v_{\rm esc}$, whereas it overestimates the escape mass of the target material when $v_{\rm imp} \gtrsim v_{\rm esc}$. To quantitatively evaluate the degree of overestimation of the escape mass of the target material, we compared HH11 to the newly derived scaling law (HG20; Equation \ref{eq_new_target}). Figure \ref{fig_comparison} shows the ratio of Equations \ref{eq_HH11} to \ref{eq_new_target}. Except for an impact angle of 15 degrees, HH11 overestimated the escape mass for an impact velocity of less than $\sim 12  v_{\rm esc}$, which exponentially increased as the impact velocity decreased. The difference became more significant for a larger impact angle towards the head-on collision. HH11 overestimated the escape mass of the target material by approximately $\sim 70$ times when $ v_{\rm imp} \sim v_{\rm esc}$ at vertical impact ($\theta=90$ degrees). For the $\theta$-averaged escape mass weighted by the $\sin(2\theta)$ distribution at HH11 and HG20, HH11 overestimated by a factor of $\sim 4$ larger than HG20.\\

\subsection{Escape and accretion by cratering impacts during planet formation}
Cratering impacts by small bodies onto a large planetary body occur much more frequently than collisions between similar-sized bodies, and the statistical distribution of the impact angle follows $\sin(2\theta)$ with a peak of $45$ degrees \citep{Sho62}. The impact velocity between two bodies depends on the degree of excitation of the system. If the system is cold and impacts consequentially occur only between local members in a radially narrow ring of bodies, the impact velocity is close to the escape velocity of the largest object in the local \citep[e.g.,][]{Ida92}. In hot systems, $-$ for example, the crossings of orbits between distant bodies \citep[e.g.,][]{Kok10} or collisions among asteroids with orbits excited by the resonance with Jupiter and Saturn \citep[e.g.,][]{Bot94} $-$ the impact velocity can be much larger than the escape velocity, depending on the orbits of two bodies.\\

\begin{figure*}
	\centering
	  \includegraphics[width=\textwidth]{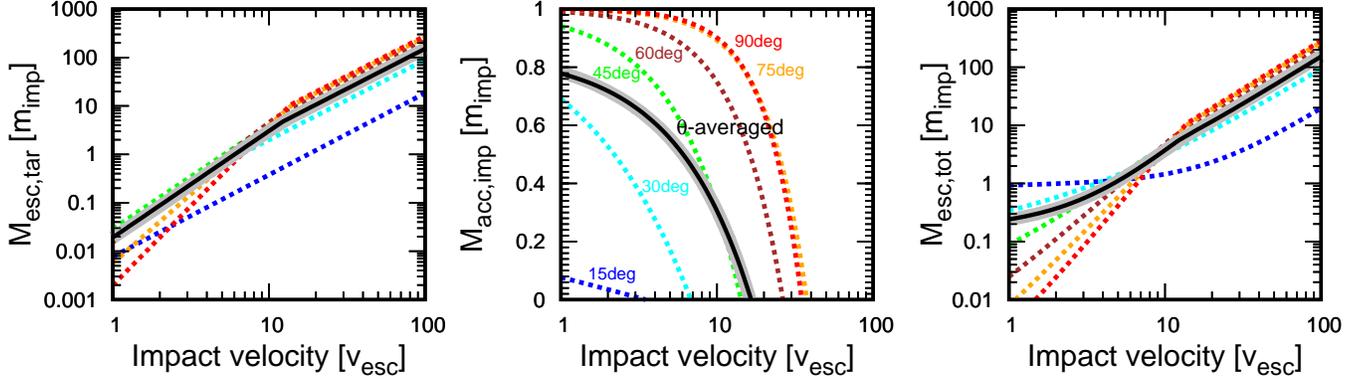}
	\caption{The escape mass of the target material (left panel; Equation \ref{eq_new_target}), the accretion mass of the impactor material onto the target surface (middle panel; Equation \ref{eq_accretion}), and the total escape mass (the escape mass of the target material + the escape mass of the impactor material; right panel) as a function of impact velocity in a unit of escape velocity. The red, orange, brown, green, cyan, and blue dashed lines represent $\theta=$90, 75, 60, 45, 30, and 15 degrees, respectively. Solid gray lines are $\theta$-averaged over $\sin(2\theta)$. Black lines are the fittings of the gray lines (equations \ref{eq_theta_average_target} and \ref{eq_theta_average_accretion} for the left and middle panels, respectively). In the right panel, the net mass escape/accretion is $M_{\rm esc/acc} = M_{\rm esc,tot} - 1$ in a unit of $m_{\rm imp}$ where a positive value indicates the net escape and vice-versa (Equation \ref{eq_erosion_accretion}).}
	\label{fig_summary}
\end{figure*}

Figure \ref{fig_summary} shows the escape mass of the target material (left panel) and the accretion mass of the impactor material onto the target (middle panel). Solid gray lines show those weighted by the $\sin(2\theta)$ distribution in Equations \ref{eq_new_target} and \ref{eq_accretion}. The $\theta$-averaged escape mass originating from the target as a function of impact velocity (left panel in Figure \ref{fig_summary}) was appropriately approximated by power law functions (solid black lines in Figure \ref{fig_summary}) as:

\begin{eqnarray}
\label{eq_theta_average_target}
	 \left< \frac{M_{\rm HG20,esc,tar}(>v_{\rm esc})}{m_{\rm imp}} \right>_{\theta} =
	 0.02 \times \left( \frac{v_{\rm imp}}{v_{\rm esc}} \right)^{2.2}
	 {\rm for} \hspace{0.5em} v_{\rm imp} \lesssim 12 v_{\rm esc}
	 \nonumber \\
	 \left< \frac{M_{\rm HG20,esc,tar}(>v_{\rm esc})}{m_{\rm imp}} \right>_{\theta} =
	 0.076 \times \left( \frac{v_{\rm imp}}{v_{\rm esc}} \right)^{1.65}
	 {\rm for} \hspace{0.5em} v_{\rm imp} \gtrsim 12 v_{\rm esc},
\end{eqnarray}

\noindent where the results are consistent with those of \cite{Sve11} reported for $v_{\rm imp} < 10 v_{\rm esc}$.\\

The accretion mass of the impactor material to the target surface is shown in the middle panel of Figure \ref{fig_summary}. As the impact velocity increases in a unit of escape velocity, the accretion mass decreases exponentially. We found that the $\theta$-averaged accretion mass from the impactor (solid gray line in Figure \ref{fig_summary}) was appropriately approximated (solid black lines in Figure \ref{fig_summary}) as follows:

\begin{equation}
\label{eq_theta_average_accretion}
	 \left< \frac{M_{\rm HG20,acc,imp}(<v_{\rm esc})}{m_{\rm imp}} \right>_{\theta} = 
	 0.85 - 0.071 \times \left( \frac{v_{\rm imp}}{v_{\rm esc}} \right)^{0.88}. 
\end{equation}
\\

The right panel of Figure \ref{fig_summary} shows the total escape mass (escape mass of the target material + escape mass of the impactor material), and a $y$-axis value of less than 1 indicates a net accretion. Thus, the net escape or accretion during a cratering impact is written as follows:

\begin{equation}
\label{eq_erosion_accretion}
	 \frac{M_{\rm esc/acc}}{m_{\rm imp}} = 
	 \frac{M_{\rm HG20,esc,tar}(>v_{\rm esc})}{m_{\rm imp}} +  \frac{M_{\rm HG20,esc,imp}(>v_{\rm esc})}{m_{\rm imp}} -1 
\end{equation}

\noindent where positive and negative values indicate net escape and accretion, respectively. As a result of the statistical impacts and $\theta$-averaged values (solid gray and black lines, respectively), net mass escape and accretion occur for impact velocities larger and smaller than $\sim 5 v_{\rm esc}$, respectively (solid gray and black lines in the right panel in Figure \ref{fig_summary}, respectively).\\

High-speed cratering impacts inevitably occur in different contexts and epochs. The solar system may have experienced a cataclysm phase such as the Nice model \citep[e.g.,][]{Gom05,Tsi05}, the "Grand-tack" hypothesis \citep[e.g.,][]{Wal11,Wal16} or the "early instability" scenario \citep[e.g.,][]{Cle18}. In these scenarios, planetesimals are gravitationally scattered by giant planets, and high-velocity collisions with terrestrial planets would take place \citep[e.g.,][]{Moj19,Bra20}. The current typical collision velocities among asteroids are $v_{\rm imp} \sim 5$ km s$^{-1}$ \citep[e.g.,][]{Bot94}, and the escape velocity of the largest asteroid, Ceres, is $v_{\rm esc} \sim 0.5$ km s$^{-1}$, indicating that the collisions among asteroids $-$ impact velocity is more than $\sim 10$ times the escape velocity $-$ are erosive, and mass escape is expected. Giant impacts, such as those that formed the Moon \citep{Bot15} and/or the Martian moons \citep{Hyo18}, would distribute impact debris throughout the inner solar system, and high-velocity collisions ($> 5$ km s$^{-1}$) between the debris and asteroids and/or planets may correspondingly occur. Such erosive impacts would play critical roles in characterizing the geomorphic and geochemical features of the surfaces of planets and asteroids.\\

\section{Summary}
\label{sec_summary}
During planet formation, the impacts of small bodies on large planetary bodies $-$ cratering impacts $-$ are inevitable and numerous events. Cratering impacts could lead to mass escape of the target material and mass accretion of the impactor material depending on the impact conditions. A fraction of impact ejecta of the target material $M_{\rm esc,tar}(>v_{\rm esc})$ escapes from the gravity of the target and becomes significant for a larger impact velocity $v_{\rm imp}$. A widely known $ M_{\rm esc,tar}(>v_{\rm esc})-v_{\rm imp}$ relationship (HH11; Equation \ref{eq_HH11}) under the point-source assumption \citep{Hol07} $-$ the size of the crater is assumed to be much larger than that of the impactor $-$ is often used in the planetary community. On the one hand, the point-source scaling law reproduces the results of impact experiments in which the ejecta velocity was much smaller than the impact velocity. On the other hand, in the case of high-speed ejecta that escapes from the target gravity, predictions by the point-source scaling law may not be appropriate, because the launch point is close to the impact point where the point-source assumption would fail. However, the quantitative limitation of the point-source solution was unclear.\\

In this study, we aimed to understand the escape mass of the target material and the accretion mass of the impactor material onto the target surface by cratering impacts. We performed an extensive number of cratering impact simulations of small bodies on a large rocky target. We explored a wide range of impact parameters: $v_{\rm imp}=6-62$ km s$^{-1}$ and $\theta=15-90$ degrees for $v_{\rm esc}=1-10$ km s$^{-1}$. We distinguished two distinct escape mass sources : target material and impactor material.\\

The numerical results of the escape mass of the target material were compared to the point-source scaling law (Equation \ref{eq_HH11}). We showed that HH11 correctly predicted the escape mass of the target material for $v_{\rm imp} \gtrsim 12   v_{\rm esc}$. However, the point-source scaling law (HH11; Equation \ref{eq_HH11}) overestimated the escape mass of the target material up to a factor of $\sim 70$ when $ v_{\rm imp} \lesssim 12 v_{\rm esc}$ (see Figure \ref{fig_comparison}). The degree of overestimation of HH11 depended on the impact angle and became more significant for smaller impact velocities and towards the vertical impact (Figure \ref{fig_comparison}).\\

In Section \ref{sec_target}, using the results of numerical simulations, we derived a new scaling law of the escape mass of the target material by a cratering impact, which can be used within and beyond the limitation of the point-source assumption (Equation \ref{eq_new_target}). We found that the same power-law dependence as HH11 ($M(>v_{\rm esc})/m_{\rm imp} \propto (v_{\rm esc}/v_{\rm imp}\sin(\theta))^{-3\mu}$) could be used beyond the limitation of the point-source assumption by correcting its exponent and coefficient as a function of the impact angle. The newly derived scaling law (Equation \ref{eq_new_target}) is applicable within a wide range of cratering impact conditions to estimate the escape mass of the target material when the escape velocity is large enough to neglect the material strength.\\

A fraction of the impactor material also escapes upon a cratering impact, and the rest of the impactor material accretes onto the target surface (Section \ref{sec_impactor}). The accretion mass originating from the impactor (accretion mass of the impactor material, $m_{\rm imp,acc}$) is given by the mass balance as $m_{\rm imp,acc} = m_{\rm imp} - m_{\rm imp,esc}$. By evaluating $m_{\rm imp,esc}$ from SPH simulations and by considering the mass balance, we estimated $m_{\rm imp,acc}$ and derived a scaling law that predicted the accretion mass of the impactor material on the target surface by using a power law function (Equations \ref{eq_accretion} and \ref{eq_theta_average_accretion}). The accretion mass of the impactor material exponentially decreased as the impact velocity increased, and almost all the mass of the impactor escaped on $\theta$-average for $v_{\rm imp}  \gtrsim 17 v_{\rm esc}$ (middle panel of Figure \ref{fig_summary}). When $v_{\rm imp} \sim v_{\rm esc}$, on $\theta$-average $\sim 80$ \% of the mass of impactor accreted on the target (Figure \ref{fig_summary}).\\

In a real system, the impact velocity depends on the degree of excitation of the system (Section \ref{sec_application}). Our newly derived scaling law (Equation \ref{eq_new_target}) indicates that net mass escape occurs for $v_{\rm imp} \gtrsim 5 v_{\rm esc}$ on $\theta$-average (right panel of Figure \ref{fig_summary}) and vice versa for the net accretion. The new scaling laws derived in this study would be useful for investigating the cumulative effect of numerous small impacts on any large planetary bodies with escape velocities large enough to neglect the material strength. Finally, the conclusions of previous studies that investigated the mass escape of large planetary bodies by using the point-source scaling law (HH11; Equation \ref{eq_HH11}) may be largely changed by redoing their work using our new scaling laws (HG20; Equations \ref{eq_new_target} and \ref{eq_accretion}). This is because most of the planetary impacts would occur with $v_{\rm imp} \lesssim 12   v_{\rm esc}$.\\

\acknowledgments
We thank Henry Jay Melosh for his constructive comments that greatly helped improve the manuscript. R.H. acknowledges the financial support of JSPS Grants-in-Aid (JP17J01269, 18K13600). H.G. acknowledges the financial support of MEXT KAKENHI Grant (JP17H06457), and JSPS Kakenhi Grant (JP17H02990 and 19H00726). We would like to thank Editage (www.editage.com) for English language editing.

\bibliography{Cratering}

\begin{thebibliography}{}
\expandafter\ifx\csname natexlab\endcsname\relax\def\natexlab#1{#1}\fi

\bibitem[{{Artemieva} \& {Shuvalov}(2008)}]{Art08}
{Artemieva}, N.~A., \& {Shuvalov}, V.~V. 2008, Solar System Research, 42, 329

\bibitem[{Asphaug(2010)}]{Asp10}
Asphaug, E. 2010, Chemie der Erde, 70, 199

\bibitem[{{Benz} \& {Asphaug}(1999)}]{Ben99}
{Benz}, W., \& {Asphaug}, E. 1999, \icarus, 142, 5

\bibitem[{Bottke {et~al.}(1994)Bottke, Nolan, Greenberg, \& Kolvoord}]{Bot94}
Bottke, W., Nolan, M., Greenberg, R., \& Kolvoord, R. 1994, Icarus, 107, 255

\bibitem[{Bottke {et~al.}(2015)Bottke, Vokrouhlick{\'{y}}, Marchi, Swindle,
  Scott, Weirich, \& Levison}]{Bot15}
Bottke, W., Vokrouhlick{\'{y}}, D., Marchi, S., {et~al.} 2015, Science, 348,
  321

\bibitem[{Brasser {et~al.}(2020)Brasser, Werner, \& Mojzsis}]{Bra20}
Brasser, R., Werner, S., \& Mojzsis, S. 2020, Icarus, 338, 113514

\bibitem[{Clement {et~al.}(2018)Clement, Kaib, Raymond, \& Walsh}]{Cle18}
Clement, M.~S., Kaib, N.~A., Raymond, S.~N., \& Walsh, K.~J. 2018, Icarus, 311,
  340

\bibitem[{{Fujiwara} {et~al.}(1977){Fujiwara}, {Kamimoto}, \&
  {Tsukamoto}}]{Fuj77}
{Fujiwara}, A., {Kamimoto}, G., \& {Tsukamoto}, A. 1977, \icarus, 31, 277

\bibitem[{{Fujiwara} \& {Tsukamoto}(1980)}]{Fuj80}
{Fujiwara}, A., \& {Tsukamoto}, A. 1980, \icarus, 44, 142

\bibitem[{Gault {et~al.}(1963)Gault, Shoemaker, Moore, Aeronautics,
  Administration, \& Center}]{Gau63}
Gault, D.~E., Shoemaker, E.~M., Moore, H.~J., {et~al.} 1963, {Spray Ejected
  from the Lunar Surface by Meteoroid Impact}, NASA technical note (National
  Aeronautics and Space Administration)

\bibitem[{Genda {et~al.}(2015)Genda, Fujita, Kobayashi, Tanaka, \& Abe}]{Gen15}
Genda, H., Fujita, T., Kobayashi, H., Tanaka, H., \& Abe, Y. 2015, Icarus, 262,
  58

\bibitem[{{Genda} {et~al.}(2017){Genda}, {Fujita}, {Kobayashi}, {Tanaka},
  {Suetsugu}, \& {Abe}}]{Gen17}
{Genda}, H., {Fujita}, T., {Kobayashi}, H., {et~al.} 2017, \icarus, 294, 234

\bibitem[{Gomes {et~al.}(2005)Gomes, Levison, Tsiganis, \& Morbidelli}]{Gom05}
Gomes, R., Levison, H., Tsiganis, K., \& Morbidelli, A. 2005, Nature, 435, 466

\bibitem[{Hartmann(1985)}]{Har85}
Hartmann, W. 1985, Science, 63, 69

\bibitem[{Hayashi {et~al.}(1985)Hayashi, Nakazawa, \& Nakagawa}]{Hay85}
Hayashi, C., Nakazawa, K., \& Nakagawa, Y. 1985, in Protostars and Planets II,
  ed. D.~Black \& M.~Matthews, 1100--1153

\bibitem[{Holsapple(1993)}]{Hol93}
Holsapple, K. 1993, Annual Review of Earth and Planetary Sciences, 21, 333

\bibitem[{Holsapple \& Housen(2007)}]{Hol07}
Holsapple, K.~A., \& Housen, K.~R. 2007, Icarus, 187, 345

\bibitem[{Holsapple \& Housen(2012)}]{Hol12}
---. 2012, Icarus, 221, 875

\bibitem[{Holsapple \& Schmidt(1987)}]{Hol87}
Holsapple, K.~A., \& Schmidt, R.~M. 1987, Journal of Geophysical Research, 92,
  6350

\bibitem[{Housen \& Holsapple(2011)}]{Hou11}
Housen, K.~R., \& Holsapple, K.~A. 2011, Icarus, 211, 856

\bibitem[{Housen {et~al.}(1983)Housen, Schmidt, \& Holsapple}]{Hou83}
Housen, K.~R., Schmidt, R.~M., \& Holsapple, K.~A. 1983, Journal of Geophysical
  Research, 88, 2485

\bibitem[{Hyodo \& Genda(2018)}]{Hyo18}
Hyodo, R., \& Genda, H. 2018, Astrophysical Journal Letters, 856,
  doi:10.3847/2041-8213/aab7f0

\bibitem[{Hyodo {et~al.}(2019)Hyodo, Kurosawa, Genda, Usui, \& Fujita}]{Hyo19}
Hyodo, R., Kurosawa, K., Genda, H., Usui, T., \& Fujita, K. 2019, Scientific
  Reports, 9, doi:10.1038/s41598-019-56139-x

\bibitem[{Ida \& Makino(1992)}]{Ida92}
Ida, S., \& Makino, J. 1992, Icarus, 98, 28

\bibitem[{{Jutzi}(2015)}]{Jut15}
{Jutzi}, M. 2015, \planss, 107, 3

\bibitem[{Kokubo \& Genda(2010)}]{Kok10}
Kokubo, E., \& Genda, H. 2010, ApJL, 714, L21

\bibitem[{Kurosawa {et~al.}(2019)Kurosawa, Genda, Hyodo, Yamagishi, Mikouchi,
  Niihara, Matsuyama, \& Fujita}]{Kur19}
Kurosawa, K., Genda, H., Hyodo, R., {et~al.} 2019, Life Sciences in Space
  Research, 23, doi:10.1016/j.lssr.2019.07.006

\bibitem[{Leinhardt \& Stewart(2012)}]{Lei12}
Leinhardt, Z.~M., \& Stewart, S.~T. 2012, ApJ, 745, 79

\bibitem[{Lucy(1977)}]{Luc77}
Lucy, L. 1977, AJ, 82, 1013

\bibitem[{Melosh(1989)}]{Mel89}
Melosh, H. 1989, {Impact cratering : a geologic process}

\bibitem[{{Melosh}(1984)}]{Mel84}
{Melosh}, H.~J. 1984, \icarus, 59, 234

\bibitem[{Michikami {et~al.}(2007)Michikami, Moriguchi, Hasegawa, \&
  Fujiwara}]{Mic07}
Michikami, T., Moriguchi, K., Hasegawa, S., \& Fujiwara, A. 2007, planss, 55,
  70

\bibitem[{Mojzsis {et~al.}(2019)Mojzsis, Brasser, Kelly, Abramov, \&
  Werner}]{Moj19}
Mojzsis, S.~J., Brasser, R., Kelly, N.~M., Abramov, O., \& Werner, S.~C. 2019,
  ApJ, 881, 44

\bibitem[{Monaghan(1992)}]{Mon92}
Monaghan, J. 1992, ARA{\&}A, 30, 543

\bibitem[{{Movshovitz} {et~al.}(2016){Movshovitz}, {Nimmo}, {Korycansky},
  {Asphaug}, \& {Owen}}]{Mov16}
{Movshovitz}, N., {Nimmo}, F., {Korycansky}, D.~G., {Asphaug}, E., \& {Owen},
  J.~M. 2016, \icarus, 275, 85

\bibitem[{Nakamura {et~al.}(1992)Nakamura, Suguiyama, \& Fujiwara}]{Nak92}
Nakamura, A., Suguiyama, K., \& Fujiwara, A. 1992, Icarus, 100, 127

\bibitem[{Okamoto {et~al.}(2020)Okamoto, Kurosawa, Genda, \& Matsui}]{Oka20}
Okamoto, T., Kurosawa, K., Genda, H., \& Matsui, T. 2020, {Impact Ejecta near
  the Impact Point Observed using Ultra-high-speed Imaging and SPH Simulations,
  and a Comparison of the Two Methods}, , , arXiv:2003.08103

\bibitem[{{Okeefe} \& {Ahrens}(1977)}]{Oke77}
{Okeefe}, J.~D., \& {Ahrens}, T.~J. 1977, Science, 198, 1249

\bibitem[{Piekutowski(1980)}]{Pie80}
Piekutowski, A. 1980, Lunar and Planetary Science Conference Proceedings, 3,
  2129

\bibitem[{Safronov(1972)}]{Saf72}
Safronov, V.~S. 1972, Israel Program for Scientific Translations

\bibitem[{Shoemaker(1962)}]{Sho62}
Shoemaker, E.~M. 1962, Physics and Astronomy of the Moon, 283

\bibitem[{{Shuvalov} \& {Artemieva}(2006)}]{Shu06}
{Shuvalov}, V.~V., \& {Artemieva}, N.~A. 2006, in 37th Annual Lunar and
  Planetary Science Conference, ed. S.~{Mackwell} \& E.~{Stansbery}, Lunar and
  Planetary Science Conference, 1168

\bibitem[{Svetsov(2011)}]{Sve11}
Svetsov, V. 2011, Icarus, 214, 316

\bibitem[{Tillotson(1962)}]{Til62}
Tillotson, J. 1962, {Metallic Equations of State For Hypervelocity Impact},
  General Atomic Report GA-3216. 1962. Technical Repor, ,

\bibitem[{Tsiganis {et~al.}(2005)Tsiganis, Gomes, Morbidelli, \&
  Levison}]{Tsi05}
Tsiganis, K., Gomes, R., Morbidelli, A., \& Levison, H. 2005, Nature, 435, 459

\bibitem[{Tsujido {et~al.}(2015)Tsujido, Arakawa, Suzuki, \& Yasui}]{Tsu15}
Tsujido, S., Arakawa, M., Suzuki, A.~I., \& Yasui, M. 2015, Icarus, 262, 79

\bibitem[{Walsh \& Levison(2016)}]{Wal16}
Walsh, K.~J., \& Levison, H.~F. 2016, AJ, 152, 68

\bibitem[{Walsh {et~al.}(2011)Walsh, Morbidelli, Raymond, O'Brien, \&
  Mandell}]{Wal11}
Walsh, K.~J., Morbidelli, A., Raymond, S.~N., O'Brien, D.~P., \& Mandell, A.~M.
  2011, Nature, 475, 206

\end{thebibliography}
\bibliographystyle{aasjournal}

\end{document}